\documentclass[dvipsnames,format=sigconf]{acmart}

\setcopyright{none}
\renewcommand\footnotetextcopyrightpermission[1]{}
\usepackage[ruled]{algorithm2e} 
\usepackage{algorithmic} 
\newcommand\algorithmicprocedure{\textbf{procedure}}
\newcommand{\algorithmicendprocedure}{\algorithmicend\ \algorithmicprocedure}
\makeatletter
\newcommand\PROCEDURE[3][default]{%
  \ALC@it
  \algorithmicprocedure\ \textsc{#2}(#3)%
  \ALC@com{#1}%
  \begin{ALC@prc}%
}
\newcommand\ENDPROCEDURE{%
  \end{ALC@prc}%
  \ifthenelse{\boolean{ALC@noend}}{}{%
    \ALC@it\algorithmicendprocedure
  }%
}
\newenvironment{ALC@prc}{\begin{ALC@g}}{\end{ALC@g}}
\makeatother

\usepackage{mathtools}
\mathtoolsset{showonlyrefs}

\usepackage{enumitem}

\newcommand\myatop[2]{\big[{\genfrac{}{}{0pt}{}{#1}{#2}}\big]} 

\AtBeginDocument{%
  \providecommand\BibTeX{{%
    \normalfont B\kern-0.5em{\scshape i\kern-0.25em b}\kern-0.8em\TeX}}}

\setcopyright{acmcopyright}
\copyrightyear{2023}
\acmYear{2023}
\acmDOI{10.1145/3583131.3590363} 

\acmConference[GECCO '23]{The Genetic and Evolutionary Computation Conference 2023}{July 15--19, 2023}{Lisbon, Portugal}

\setcopyright{acmlicensed}\acmConference[GECCO '23]{Genetic and Evolutionary Computation Conference}{July 15--19, 2023}{Lisbon, Portugal}
\acmBooktitle{Genetic and Evolutionary Computation Conference (GECCO '23), July 15--19, 2023, Lisbon, Portugal}
\acmPrice{15.00}
\acmISBN{979-8-4007-0119-1/23/07} 

\begin{document}

\title{Positive definite nonparametric regression using an evolutionary algorithm with application to covariance function estimation}

\author{Myeongjong Kang}
\affiliation{%
  \institution{Texas A{\&}M University}
  \city{College Station}
  \country{TX, USA}}
\email{myeongjong@tamu.edu}

\renewcommand{\shortauthors}{Kang}

\begin{abstract}
We propose a novel nonparametric regression framework subject to the positive definiteness constraint. It offers a highly modular approach for estimating covariance functions of stationary processes. Our method can impose positive definiteness, as well as isotropy and monotonicity, on the estimators, and its hyperparameters can be decided using cross validation. We define our estimators by taking integral transforms of kernel-based distribution surrogates. We then use the iterated density estimation evolutionary algorithm, a variant of estimation of distribution algorithms, to fit the estimators. We also extend our method to estimate covariance functions for point-referenced data. Compared to alternative approaches, our method provides more reliable estimates for long-range dependence. Several numerical studies are performed to demonstrate the efficacy and performance of our method. Also, we illustrate our method using precipitation data from the Spatial Interpolation Comparison 97 project.
\end{abstract}



\maketitle

\settopmatter{printacmref=false}
\setcopyright{none}
\renewcommand\footnotetextcopyrightpermission[1]{}
\pagestyle{plain}


\section{Introduction \label{sec:intro}}

There have been numerous studies on nonparametric regression analysis under shape constraints, including monotonicity and convexity \citep{groeneboom2014nonparametric,guntuboyina2018nonparametric}. Nonparametric regression techniques are typically considered as solutions to overcome an inherent weakness of parametric regression models. The failure of the parametric approaches arises when the models are so restricted that the relationship between explanatory and response variables cannot be described properly. Since the nonparametric approaches only assume that the models belong to a certain infinite dimensional collection of functions, they are more flexible in estimating the target function \citep{hardle1990applied,eubank1999nonparametric}. However, a number of regression problems require estimators to have certain features or shapes. For instance, to estimate covariance functions or variograms using regression techniques, positive definiteness has been considered as a crucial constraint since the covariance estimates are ultimately used for kriging or prediction \citep{cressie2015statistics,gorsich2000variogram}. Also, isotropy and monotonicity are typically regarded as reasonable assumptions in practice. To tackle this issue, the most widely used approach is to use pre-selected parametric covariance functions such as Mat{\'e}rn or squared exponential kernels \citep{genton2001classes}, since positive definiteness has not received abundant attention in nonparametric estimation literature, compared to other shape constraints.

There have been several studies on positive definite nonparametric regression approaches. After \cite{armstrong1981variogram} emphasized the importance of using positive definite functions for estimating covariance functions, \cite{christakos1984problem} proposed several criteria that allow to check whether a function is positive definite. Although they focused on parametric covariance function estimation, they provided necessary and sufficient conditions of ``permissible'' covariance function models. Notably, \cite{shapiro1991variogram} proposed a nonparametric method to estimate isotropic covariance functions using non-negative regression techniques, which is referred to as the Shapiro-Botha method in this paper. Also, \cite{hall1994nonparametric} suggested a different nonparametric method using kernel regression techniques, which is here referred to as the Hall-Fisher-Hoffmann method. Since both methods provided promising ideas, there have been several related studies: For instance, the Shapiro-Botha method was evaluated against typical parametric covariance models by \cite{cherry1996evaluation}. \cite{garcia2003local} and \cite{garcia2004nonparametric} proposed nonparametric variogram estimators based on popular nonparametric regression techniques. Note that they used the Shapiro-Botha method to guarantee positive definiteness of the nonparametric estimators. In addition, \cite{garcia2007asymptotic} discussed the asymptotic properties of a Nadaraya-Watson-type variogram estimator. \cite{gorsich2000variogram} introduced a procedure for selecting a parametric variogram model based on derivatives of the Shapiro-Botha estimator. A numerical experiment for comparing the estimators was provided in \cite{menezes2005comparison}. \cite{fernandez2014nonparametric} proposed a bias-corrected version of the Shapiro-Botha estimator. 

Both the Shapiro-Botha and Hall-Fisher-Hoffmann methods have advantages and disadvantages. The Shapiro-Botha method is computationally inexpensive and can be easily implemented. Also, one can impose additional assumptions, such as smoothness or monotonicity. The Hall-Fisher-Hoffmann method has nice large-sample properties. However, one disadvantage of using the Shapiro-Botha method is that there is no clear rule for node selection. Both \cite{shapiro1991variogram} and \cite{cherry1996evaluation} considered several ways to select the nodes, for instance, 200 equidistant nodes. But, according to the simulation results in \cite{shapiro1991variogram} and \cite{cherry1996evaluation}, the use of equidistant nodes causes spurious oscillations of the estimator. To avoid this issue, \cite{genton2002nonparametric} claimed that the nodes should be chosen as zeros of Bessel functions. The node selection, however, remains a challenging task for using the Shapiro-Botha estimator. On the other hand, the Hall-Fisher-Hoffmann method requires deciding on a proper positive rendering operator for coercing the Fourier transform of kernel-type covariance function estimators into non-negative functions. Arguably, the positive rendering operator selection is also a difficult task, since it is unclear whether positive rendering affects the finite sample properties of the estimator. Furthermore, the Hall-Fisher-Hoffmann method only considers one-dimensional cases and is therefore not applicable to spatial data analysis. Note that several other approaches also consider only one-dimensional cases \citep[e.g.][]{cai2010nonparametric}. To the best of our understanding, there is no widely used nonparametric approach for covariance function estimation.

In this paper, we propose a new nonparametric regression framework subject to the positive definiteness constraint, which can be useful for covariance function estimation. It provides a highly modular approach for estimating covariance functions of stationary processes. We introduce three different estimators: Positive definite, isotropic positive definite, and monotone isotropic positive definite estimators. Like the other two popular methods, our approach is based on Bochner's theorem \citep{bochner1933integration} which implies that a continuous positive definite function can be formulated as the Fourier transform of a finite non-negative Borel measure. But, unlike the other two methods, our method allows to decide hyperparameters (e.g., bandwidth of kernel and size of pseudo datasets), for example, using cross validation. To obtain the advantages, the estimators are defined by taking integral transforms of kernel-based distribution surrogates. We suggest using the iterated density estimation evolutionary algorithm which is a variant of estimation of distribution algorithms \citep[EDAs;][]{larranaga2002review}. Our numerical experiments show that the proposed method can be more accurate and stable than the Shapiro-Botha method, which implies that the estimators can be useful for kriging. The code needed to reproduce our results can be found at \url{https://github.com/myeongjong/NPcov}.

The rest of this paper is organized as follows. Our method is presented in Section~\ref{sec:method}. Our three different estimators are presented in Section~\ref{sec:model}. We provide several closed-form examples of the estimators in Section~\ref{sec:example}. We propose to obtain the estimators using the algorithm presented in Section~\ref{sec:estim}. We discuss how to estimate covariance functions using our method in Section~\ref{sec:covEst}. We evaluate our estimators on synthetic data in Section~\ref{sec:simul}. In specific, we consider a typical regression problem in Section~\ref{sec:simReg} and covariance function estimation in Section~\ref{sec:simCov}. We also illustrate our method with real data from the Spatial Interpolation Comparison (SIC) 97 project. Section~\ref{sec:conc} concludes and discusses future work. Appendices contain proofs, additional numerical results, and empirical evidence on the convergence of the iterated density estimation evolutionary algorithm we use in this paper.


\section{Method \label{sec:method}}

In this section, we propose a positive definite nonparametric regression framework using kernel smoothing techniques \citep{silverman1986density,wand1994kernel}. Our estimators are based on Bochner's theorem \citep{bochner1933integration} and an inference algorithm is obtained by adopting the iterated density estimation evolutionary algorithm \citep[IDEA;][]{bosman1999algorithmic}. 


\subsection{Model and estimation \label{sec:model}}

Suppose that data $(\mathbf{x}_1, y_1), \cdots , (\mathbf{x}_n, y_n)$ are observed, where $\mathbf{x}_i \in \mathbb{R}^d$, $d \ge 1$, and $y_i \in \mathbb{R}$ are the $i$th explanatory and response variables, respectively. Also, assume that the response can be modeled with the unknown regression function $f$, which implies that
\begin{equation}
y_i = f(\mathbf{x}_i ) + \epsilon_i \, , \quad i = 1, 2, \cdots, n,
\end{equation}
where $\epsilon_i$ is the $i$th observation error. Our goal here is to estimate the regression function $f$ under reasonable assumptions. We consider three different types of the regression function: ($i$) In the general case, we assume that $f$ is continuous and positive definite. Also, for the sake of simplicity, we assume that $f(\mathbf{0}) = 1$. ($ii$) In the isotropic case, it is additionally assumed in the general case that $f$ is isotropic. ($iii$) In the monotone case, it is additionally assumed in the isotropic case that $f$ is monotone.

For the general case, Bochner's theorem \citep{bochner1933integration} says that $f$ can be represented in the form
\begin{equation}
\label{eq:regfun1}
f(\mathbf{x})  = \int_{\mathbb{R}^d} e^{-2 \pi i \mathbf{x} \cdot \mathbf{u}}\ {\rm d}F(\mathbf{u}),
\end{equation}
where $F$ is a distribution function on $\mathbb{R}^d$. It is clear that an estimator of $f$ can be obtained by imposing a restriction on $F$. In this paper, we define a kernel-based surrogate of $F$ by
\begin{equation}
\tilde{F}_m (\mathbf{u}) = \frac{1}{m} \sum_{i=1}^{m} \int_{-\infty}^{\mathbf{u}}  \mathbf{K}_{\mathbf{H}} (\tilde{\mathbf{u}} - \mathbf{v}_i ) {\rm d}\tilde{\mathbf{u}},
\end{equation} 
where $m \ge 1$, $\mathbf{v}_1 , \cdots , \mathbf{v}_m \in \mathbb{R}^d$, $\mathbf{K}$ is a $d$-variate kernel function satisfying $\int \mathbf{K}(\mathbf{t}) {\rm d}\mathbf{t} = 1$, $\mathbf{H}$ is a positive definite $d \times d$ matrix called the bandwidth matrix, and $\mathbf{K}_{\mathbf{H}} (\mathbf{t}) = |\mathbf{H}|^{-\frac{1}{2}} \mathbf{K} (\mathbf{ H}^{-\frac{1}{2}} \mathbf{t})$. See \cite{wand1994kernel} for details about the formulation. We should mention that that although $\mathbf{v}_1 , \cdots , \mathbf{v}_m$ are typically observed as ``data'', this is not the case; they are vectors specifying an estimator $\tilde{F}_m$. We call $\mathbf{v}_1 , \cdots , \mathbf{v}_m$ pseudo data. Also, we call $m$ the size of pseudo data. Therefore, an estimator of $f$ can be modeled by replacing $F$ in Equation~\eqref{eq:regfun1} by $\tilde{F}_m$ as follows:
\begin{equation}
\label{eq:estfun1}
\hat{f}_m (\mathbf{x})
= \int_{\mathbb{R}^d} e^{-2 \pi i \mathbf{x} \cdot \mathbf{u}}\ {\rm d}\tilde{F}_m (\mathbf{u}) 
= \frac{1}{m} \sum_{i=1}^{m} \cos (2 \pi \mathbf{x} \cdot \mathbf{v}_i)\mathcal{F}(\mathbf{K}_{\mathbf{H}})(\mathbf{x}),
\end{equation}
where $\mathcal{F}(\mathbf{K}_{\mathbf{H}})$ is the Fourier transform of $\mathbf{K}_{\mathbf{H}}$. This estimator is referred to as the general positive definite regression estimator or simply general estimator. Note that one can use any popular kernel functions including uniform, Epanechnikov and Gaussian kernels.

Although the general estimator $\hat{f}_m$ is formulated with the commonly used functions and operators, it can be computationally expensive since it is necessary to decide hyperparameters $m$ and $\mathbf{H}$ and optimize the pseudo data $\mathbf{v}_1 , \cdots , \mathbf{v}_m$ as well. This can be achieved through the algorithm we propose in Section~\ref{sec:estim}, but the computational issue can be avoided by making an additional assumption on the shape of the regression function; in this paper, we consider isotropy and monotonicity. A main reason we consider isotropic functions is that isotropy is one of the most frequently assumed properties for modeling covariance functions. Furthermore, in practice, it is typically assumed that covariance functions are not only isotropic but also monotonically decreasing.

In the isotropic case that there is a function $g : [0, \infty) \rightarrow \mathbb{R}$ such that $f(\mathbf{x}) = g(|\mathbf{x}|)$ for all $\mathbf{x} \in \mathbb{R}^d$, our goal is to estimate the one dimensional function $g$. Note that $g$ is continuous, positive definite and isotropic. Also, $g(0) = 1$. To obtain an estimator, we use Schoenberg's theorem \citep{schoenberg1938metric} which can be viewed as a modified version of Bochner's theorem for isotropic functions \citep{fasshauer2007meshfree,cheney2009course}. By Schoenberg's theorem, $g$ can be represented as
\begin{equation}
\label{eq:regfun2}
g(r) = \int_{0}^{\infty} \Omega_d (ru) {\rm d} G(u),
\end{equation}
where $G$ is a distribution function on $\mathbb{R}^{+}_{0} = [0, \infty)$,
\begin{equation}
\Omega_d (r) =
\left\{
	\begin{array}{ll}
		\cos(r)  & \mbox{for } d = 1 \\
		\boldsymbol{\Gamma} \left( \frac{d}{2} \right)\ \left( \frac{2}{r}\right)^{(d-2)/2} \mathbf{J}_{(d-2)/2}(r) & \mbox{for } d \ge 2
	\end{array}
\right. ,
\end{equation}
$\boldsymbol{\Gamma}$ is the gamma function and $\mathbf{J}_{\nu}$ is the Bessel function of the first kind of order $\nu \in \mathbb{R}$. Similar to the general case, we can replace $G$ using kernel smoothing. There is extensive literature on kernel density estimation with a bounded support \citep[see][for details]{zhang1999improved,karunamuni2005boundary}. Here we use the reflection method \citep{schuster1985incorporating,silverman1986density,cline1991kernel} due to the fact that one can easily formulate the model and use popular kernel functions with the reflection method. We define a surrogate of $G$ by
\begin{equation}
\label{eq:estG}
\tilde{G}_m (u) = \frac{1}{m}\sum_{i=1}^{m} \int_{0}^{u}[ K_h (\tilde{u}-v_i) + K_h (\tilde{u}+v_i)]{\rm d}\tilde{u}\ ,
\end{equation}
where $m \ge 1$, $v_1 , \cdots , v_m \in \mathbb{R}^{+}_{0}$, $K$ is a univariate kernel function satisfying $\int K(u) {\rm d}u = 1$, $h > 0$ is the bandwidth, and $K_h (t) = \frac{1}{h}K(\frac{t}{h})$. Therefore, an estimator of $g$ can be obtained by replacing $G$ in Equation~\eqref{eq:regfun2} by $\tilde{G}_m$ as follows:
\begin{multline}
\label{eq:estfun2}
\hat{g}_m (r) = \frac{1}{m} \sum_{i=1}^{m} \bigg[ \int_{0}^{\infty} \Omega_{d}(ru) K_h (u-v_i ) {\rm d}u \\+ \int_{0}^{\infty} \Omega_{d}(ru) K_h (u+v_i ) {\rm d}u\bigg].
\end{multline}
We call this estimator the isotropic positive definite regression estimator or simply isotropic estimator. Note that it is required to decide both $m$ and $h$ and also optimize pseudo data $v_1 , \cdots , v_m$. 

For the monotone case, we derive the estimator based on the approach presented by \cite{cherry1996evaluation} in the following manner: From the fact that continuous positive definite isotropic functions on $\mathbb{R}^s$ for all $s \ge 1$ are monotone, \cite{cherry1996evaluation} proposed to use $\exp (-r^2 u^2)$ instead of $\Omega_{d}(ru)$ in Equation~\eqref{eq:regfun2}. One can easily see that adopting this idea to Equation~\eqref{eq:estfun2} guarantees monotonicity of the estimator. We propose to estimate the monotone regression function by using
\begin{multline}
\label{eq:estfun3}
\hat{q}_m (r) = \frac{1}{m} \sum_{i=1}^{m} \bigg[ \int_{0}^{\infty} e^{-r^2 u^2} K_h \left(u-v_i \right) {\rm d}u \\+ \int_{0}^{\infty} e^{-r^2 u^2} K_h \left(u+v_i \right) {\rm d}u\bigg].
\end{multline}
This model is referred to as the monotone isotropic positive definite regression estimator or simply monotone estimator. Note that the monotone estimator requires to decide $m$ and $h$ and also optimize pseudo data $v_1 , \ldots , v_m$, similar to the isotropic estimator. 


\subsection{Closed-form examples \label{sec:example}}

We here obtain closed-form expressions of the estimators for several scenarios. According to our numerical experiments, although the isotropic and monotone estimators require evaluation of indefinite integrals, one can compute them numerically and also efficiently. But, it is also true that the estimators can be explicitly formulated and computed quickly by using proper kernel functions. 
We here consider three popular kernel functions: The uniform kernel is popular due to its simplicity; the Epanechnikov kernel is also popular mainly due to its theoretical properties; and the Gaussian kernel is widely used for its numerical stability and differentiability. We henceforth restrict ourselves to the cases where regression functions are isotropic. Furthermore, we assume that the dimension of the input domain $d=2$. Note that the definitions of the kernels and proofs for the following examples are in Appendix~\ref{app:proofs}.

\paragraph{Uniform kernel} Assume $r > 0$. Let $K$ be the uniform kernel. Then the isotropic estimator can be represented as
\begin{equation}
\hat{g}_m (r) = \frac{1}{2mhr} \sum_{j=1}^{m} \left[ \Lambda_1 (r(v_j +h)) - \Lambda_1 (r(v_j - h)) \right],
\end{equation}
where $\mathcal{H}_{\nu}$ is the Sturve function of order $\nu$ and $\Lambda_1$ is defined as
\begin{equation}
\Lambda_1 (r) = r \mathbf{J}_0 (r) + \frac{\pi r}{2} \left[ \mathbf{J}_1 (r) \mathcal{H}_0 (r) - \mathbf{J}_0 (r) \mathcal{H}_1 (r) \right].
\end{equation}
Also, the monotone estimator can be represented as
\begin{equation}
\hat{q}_m (r) = \frac{\sqrt{\pi}}{2mhr} \sum_{j = 1}^{m} \left[ \Phi \left(\sqrt{2}r(v_j +h)\right) - \Phi \left(\sqrt{2}r(v_j -h)\right) \right],
\end{equation}
where $\Phi$ is the distribution function of standard normal distribution.

\paragraph{Epanechnikov kernel} Assume $r > 0$. Let $K$ be the Epanechnikov kernel. Then the isotropic estimator can be represented as
\begin{multline}
\hat{g}_m (r) = \frac{3}{4mhr}\sum_{j=1}^{m}\left[ \left( 1-\frac{v_j ^2}{h^2}\right) \left( \Lambda_1 (r (v_j + h)) - \Lambda_1 (r (v_j - h))\right) \right. \\ \left. - \left( 1-\frac{v_j ^2}{h^2} \right) \left( \mathbf{J}_1 (r(v_j + h)) - \mathbf{J}_1 (r(v_j - h)) \right) \right. \\ \left. + \left( \frac{1}{r^2 h^2}\right) \left( \Lambda_0 (r (v_j + h)) - \Lambda_0 (r (v_j - h))\right) \right],
\end{multline}
where 
\begin{equation*}
\Lambda_0 (r) = \frac{\pi r}{2} \left[ \mathbf{J}_1 (r) \mathcal{H}_0 (r) - \mathbf{J}_0 (r) \mathcal{H}_1 (r) \right].
\end{equation*}
Also, the monotone estimator can be represented as
{\small
\begin{multline}
\hat{q}_m (r) = \frac{3}{4mhr} \sum_{j=1}^{m}  \Bigg[ \sqrt{\pi} \bigg( 1 - \frac{v_j ^2}{h^2} - \frac{1}{2r^2 h^2}\bigg) \Big( \Phi (r (v_j + h)) - \Phi (r (v_j - h)) \Big) \\ + \bigg( \frac{1}{r^2 h^2}\bigg) \Big( (v_j + h) e^{-r^2 (v_j - h)^2} - (v_j - h)e^{-r^2 (v_j + h)^2} \Big) \Bigg].
\end{multline}
}

\begin{figure*}[h]
    \centering
    \includegraphics[width=\textwidth]{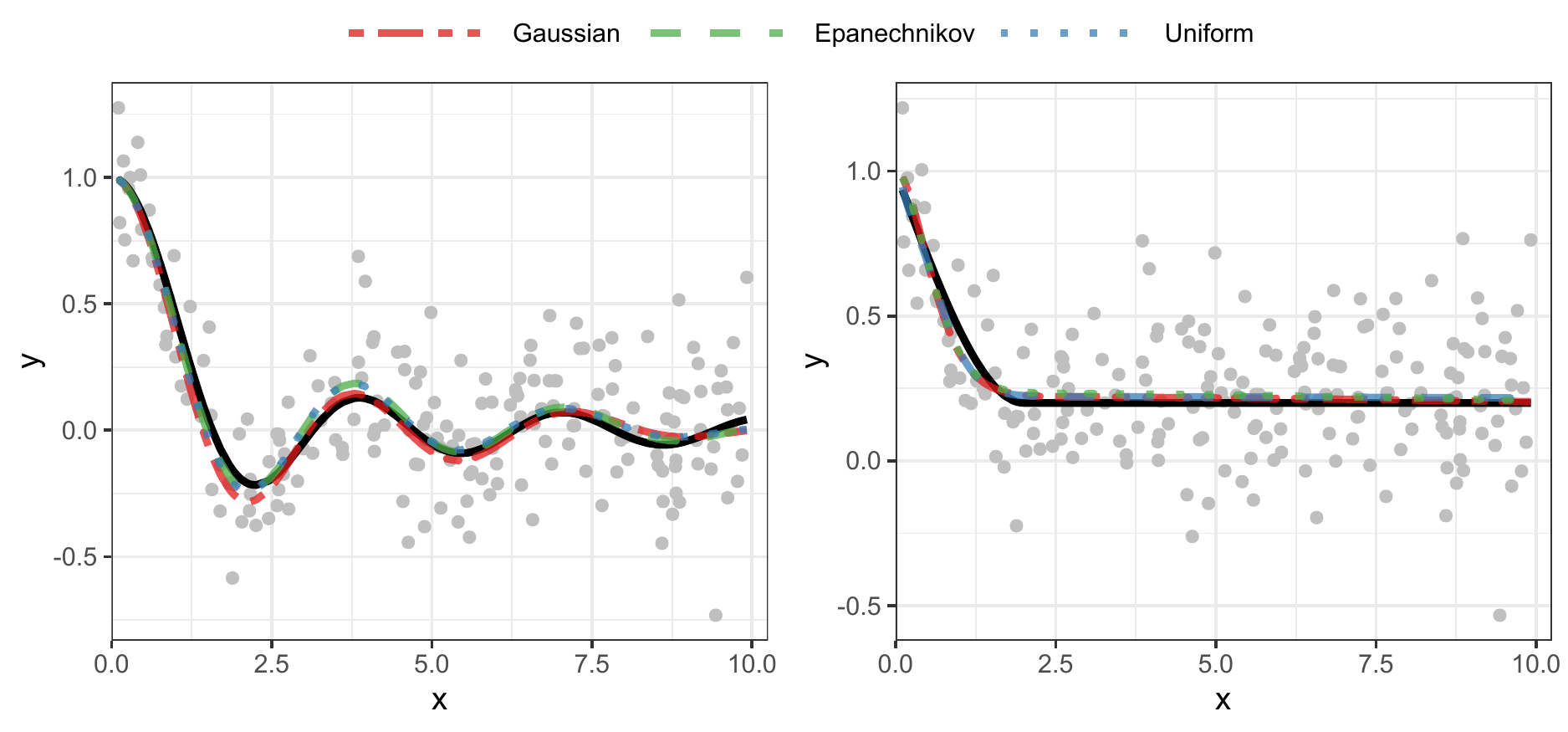}
    \caption{Comparison of three popular kernels for the isotropic (left) and monotone (right) estimators. We consider Gaussian, Epanechnikov and Uniform kernels. The regression equations~\eqref{eq:regeqn1} and~\eqref{eq:regeqn2} are considered on the left and right panels, respectively. The black lines are the true regression equations, and gray points are noisy observations. The estimators are fitted using the iterated density estimation evolutionary algorithm. Bandwidth $h$ and pseudo-data size $m$ are selected using 5-fold cross validation.}
    \label{fig:simulation_kernels}
\end{figure*}

\paragraph{Gaussian kernel} Assume $r > 0$. Let $K$ be the Gaussian kernel. Then the isotropic estimator can be represented as
{\small
\begin{align}
\hat{g}_m (r) &= \frac{1}{m} e^{-\frac{1}{4} h^2 r^2} \sum_{j=1}^{m} \mathbf{J}_0 ^{(2)} \left(\frac{\mathrm{i}\mkern1mu}{4}h^2 r^2 , rv_j , {\mathrm{i}\mkern1mu}^{-1}\right) \\
&= \frac{1}{m\pi}e^{-\frac{1}{4} h^2 r^2} \sum_{j=1}^{m}\int_{0}^{\pi} \exp \left( \frac{1}{4}h^2 r^2 \cos(2\phi) \right) \cos(rv_j \sin (\phi)) {\rm d} \phi ,
\end{align}
}
where ${\mathrm{i}\mkern1mu} = \sqrt{-1}$ and $\mathbf{J}_0 ^{(2)}$ is the $\myatop{(2)}{0}$ order generalized Bessel function \citep{dattoli1993advances}. Also, the monotone estimator can be represented as
\begin{equation}
\hat{q}_m (r) = \frac{1}{mh} \sum_{j=1}^{m} \frac{1}{\sqrt{2A(r, h)}} e^{C_{j}(r,h) },
\end{equation}
where $A(r, h) = r^2 + \frac{1}{2h^2}$ and $C_{j}(r,h)= \frac{v_j ^2}{4 h^4 A(r, h) }- \frac{v_j ^2}{2h^2}$ for $j = 1, 2, \cdots , m$. Our numerical experiments show that the estimators with the Gaussian kernel can be obtained more efficiently and differences in estimation performance between the kernels are negligible, as illustrated in Figure~\ref{fig:simulation_kernels}. Therefore, from now on we only consider the Gaussian kernel.

\subsection{Algorithm \label{sec:estim}}

To search for proper pseudo data, we suggest to use the iterated density estimation evolutionary algorithm \citep[IDEA;][]{bosman1999algorithmic}, a variant of estimation of distribution algorithms \citep[EDA;][]{larranaga2002review}, mainly because optimizing pseudo data is a non-convex problem. EDAs are probabilistic model fitting genetic algorithms, and their convergence is discussed by \cite{zhang2004convergence} and \cite{rastegar2005study}. The concepts are already introduced to statisticians by \cite{chatterjee1996genetic}. Here we focus on introducing an algorithm for obtaining isotropic and monotone estimators.

An unified framework for IDEA is specified in Algorithm~\ref{alg:IDEA}. To use the framework, one needs to first decide not only kernel-specifying hyperparameters (bandwidth of kernel $h > 0$ and size of pseudo data $m \ge 0$) but also algorithm-specifying hyperparameters (size of collection of pseudo datasets $l \ge 1$ and acceptance rate $\tau > 0$). Fortunately, it appears that $l / m$ and $\tau$ do not significantly affect the estimation performance. In numerical experiments, we assume $l = 10 m$ and $\tau = 0.1$. We use cross validation to decide kernel-specifying hyperparameters. Note that we provide numerical evidence of convergence of the IDEA in Appendix~\ref{app:conv}.

The IDEA framework requires specifying six steps: Initialization, evaluation, selection, estimation, replacement and termination. The initialization step is to generate $l$ pseudo datasets randomly from the initialized surrogate $\tilde{G}_m$ of the distribution $G$. Our numerical studies show that the exponential distribution with rate parameter 1 works properly in any case we consider in this paper. 

The evaluation step is to compute an objective function value for each pseudo dataset. Arguably, there are a variety of objective functions we can use for the algorithm. For instance, if the observation errors are independent and homoskedastic, we suggest to use the mean of squared errors (MSE): In specific, the objective function $L$ of $i$th pseudo-dataset $\Theta_{i}^{(k)}$ of $k$th step is defined as 
\begin{equation}
L(\Theta_{i}^{(k)}) = \frac{1}{n} \sum_{j=1}^{n} \left( y_j - \hat{g}_m \left(|\mathbf{x}_j|; \Theta_{i}^{(k)}\right) \right)^2
\end{equation}
where $\hat{g}_m (|\mathbf{x}_j|; \Theta_{i}^{(k)})$ is defined by Equation~\eqref{eq:estfun2},
$$
\Theta_{i}^{(k)} = \lbrace v_{1,i} ^{(k)} , \cdots , v_{m,i} ^{(k)} \rbrace    
$$
and $v_{j,i} ^{(k)}$ is the $j$th element of $i$th pseudo dataset of $k$th step for all $1 \le j \le m$, $1 \le i \le l$. But, the errors are clearly correlated and heteroscedastic for covariance function estimation scenarios. We will discuss this issue in Section~\ref{sec:covEst}. 

The selection step is to select $\lfloor \tau l \rfloor$ best pseudo datasets generated from the previous step with respect to the values of the objective function $L$. This action of $k$-step can be described as follows:
 \begin{enumerate}[leftmargin=0.75cm]
	\item[$\cdot$] Rearrange $l$ pseudo datasets $\Theta_{1}^{(k)}, \cdots , \Theta_{l}^{(k)}$ in ascending order of objective function values $L(\Theta_{1}^{(k)}), \cdots , L(\Theta_{l}^{(k)})$ and denote the ordered pseudo datasets by $\Theta_{(1)}^{(k)}, \cdots , \Theta_{(l)}^{(k)}$.	
	\item[$\cdot$] Choose $\lfloor \tau l \rfloor$ pseudo datasets $\Theta_{(1)}^{(k)}, \cdots , \Theta_{(\lfloor \tau l \rfloor)}^{(k)}$ of which objective function values are smaller than those of other pseudo datasets $\Theta_{(\lfloor \tau l \rfloor +1)}^{(k)} , \cdots , \Theta_{(l)}^{(k)}$.
\end{enumerate}

The estimation step is to estimate $G$ with the $\lfloor \tau l \rfloor$ selected pseudo datasets. In the $k$th step, one can obtain the $k$-step distribution surrogate $\tilde{G}_{m}^{(k)}$ using Equation~\eqref{eq:estG} with the merged pseudo dataset $\Theta^{(k)}$ defined by
$
\Theta^{(k)} = \Theta_{(1)}^{(k)} \cup \cdots \cup \Theta_{(\lfloor \tau l \rfloor)}^{(k)}.
$

\begin{algorithm}
\caption{Iterated density estimation evolutionary algorithm (IDEA). Note that objective function and pseudo-dataset are called OBF and PSD, respectively.}\label{alg:IDEA}
\begin{algorithmic}[1]
\PROCEDURE{IDEA}{$h$, $m$, $l$, $\tau$}
\STATE Generate a collection of $l$ PDs \COMMENT{Initialization}
\WHILE{the termination rule is true}
\STATE Evaluate OBF values of PSDs \COMMENT{Evaluation}
\STATE Select $\lfloor \tau l \rfloor$ best PSDs \COMMENT{Selection}
\STATE Estimate $G$ with the selected PSDs \COMMENT{Estimation}
\STATE Check the termination rule \COMMENT{Termination}
\IF {the termination rule is false}
\STATE Generate $l - \lfloor \tau l \rfloor$ PSDs \COMMENT{Replacement}
\ENDIF
\ENDWHILE
\STATE \textbf{return} a collection of $\lfloor \tau l \rfloor$ selected PSDs
\ENDPROCEDURE
\end{algorithmic}
\end{algorithm}

The replacement step is to replace $l - \lfloor \tau l \rfloor$ worst pseudo datasets contained in the collection of pseudo datasets with new pseudo datasets. This can be accomplished using the simulation approach proposed by \cite{silverman1986density}. Replacement of $k$th step can be described as follows:
\begin{itemize}[leftmargin=0.75cm]
	\item[$\cdot$] Denote the merged pseudo dataset $\Theta^{(k)}$ by $\lbrace \tilde{v}_1 , \cdots , \tilde{v}_{m\lfloor \tau l \rfloor} \rbrace$. 	
	\item[$\cdot$] Choose $I_1 , \cdots , I_{m(l-\lfloor \tau l \rfloor )}$ uniformly with replacement from $\lbrace 1,\cdots , m \lfloor \tau l \rfloor \rbrace$.
	\item[$\cdot$] Draw random samples $e_1 , \cdots , e_{m(l-\lfloor \tau l \rfloor )}$ from the kernel $K$.
	\item[$\cdot$] Set $\tilde{v}_{i}' = \mid \tilde{v}_{I_i} + e_i \mid$ for all $i = 1, 2, \cdots , m(l-\lfloor \tau l \rfloor )$.
	\item[$\cdot$] Divide $\lbrace \tilde{v}_1 ', \cdots , \tilde{v}_{m(l-\lfloor \tau l \rfloor )}' \rbrace$ into $l-\lfloor \tau l \rfloor$ pseudo datasets \\ $\Theta_{\lfloor \tau l \rfloor +1}^{(k+1)}, \cdots , \Theta_{l}^{(k+1)}$ uniformly. 
	\item[$\cdot$] Set $\Theta_{1}^{(k+1)} = \Theta_{(1)}^{(k)}, \cdots , \Theta_{\lfloor \tau l \rfloor}^{(k+1)} = \Theta_{(\lfloor \tau l \rfloor)}^{(k)}$.
\end{itemize}
Then, $\Theta_{1}^{(k+1)}, \ldots, \Theta_{l}^{(k+1)}$ are the $(k+1)$-step pseudo datasets of Algorithm~\ref{alg:IDEA}. The termination step is based on the concept of Kullback-Leibler (KL) divergence \citep{kullback1951information}. Let $\tilde{\psi}^{(k)}$ denote the kernel density surrogate corresponding to $\tilde{G}_{m}^{(k)}$. Then, KL divergence from $\tilde{G}_{m}^{(k-1)}$ to $\tilde{G}_{m}^{(k)}$ can be computed by
\begin{equation}
D_{KL} (\tilde{G}_{m}^{(k)}, \tilde{G}_{m}^{(k-1)}) = \frac{1}{m \lfloor \tau l \rfloor} \sum_{i=1}^{\lfloor \tau l \rfloor}\sum_{j=1}^{m} \log \left( \frac{\hat{\psi}_{m}^{(k)} (v_{j,i} ^{(k-1)})}{\hat{\psi}_{m}^{(k-1)} (v_{j,i} ^{(k-1)})} \right).
\end{equation}
We terminate Algorithm~\ref{alg:IDEA} when $D_{KL} < 10^{-3}$ has been met five times consecutively for stability. Finally, we obtain the estimator using Equation~\eqref{eq:estfun2} with the pseudo data obtained by merging a collection of $\lfloor \tau l \rfloor$ selected pseudo-datasets at the last step of Algorithm~\ref{alg:IDEA}.


\subsection{Application to covariance function estimation \label{sec:covEst}}

In this section, we propose a covariance function estimation procedure using our nonparametric positive definite regression estimators. Consider a stationary process $\left\lbrace Z(\mathbf{s}) \mid \mathbf{s} \in D \right\rbrace$, where the domain $D$ is a bounded subset of $\mathbb{R}^d$. The covariance function $C:\mathbb{R}^d \rightarrow \mathbb{R}$ can be defined as 
\begin{equation}
C(\mathbf{s}-\mathbf{s}') = {\rm cov} \left( Z(\mathbf{s}), Z(\mathbf{s}') \right) \quad \text{for all}\  \mathbf{s} , \mathbf{s}' \in D.
\end{equation}
Let $Z(\mathbf{s}_1), Z(\mathbf{s}_2), \cdots, Z(\mathbf{s}_{w})$ denote a realization of the stationary process $Z$. Then, one can obtain point estimates of the covariance function using the method proposed by \cite{matheron1962traite}:
\begin{equation}
\hat{c}_{i,j} = \hat{C}(\mathbf{s}_i-\mathbf{s}_j) = \left( Z(\mathbf{s}_i) - \bar{Z}\right) \left( Z(\mathbf{s}_j) - \bar{Z}\right)
\end{equation}
where $\bar{Z} = \frac{1}{w} \sum_{i=1}^{w} Z(\mathbf{s}_i)$. According to \cite{hall1994nonparametric}, the set of observations may either include or exclude the diagonal terms corresponding to $i=j$. Here, we use diagonal terms only for initialization. This can be framed as a heteroscedastic regression problem with observations $\left( \mathbf{r}_{i,j}, \hat{c}_{i,j}\right)$ where $\mathbf{r}_{i, j} = \mathbf{s}_i- \mathbf{ s}_j$. The presence of heteroscedasticity is induced by the fact that the classical point estimates are highly correlated and variable. To tackle this issue, \cite{journel1978mining} recommended to take large enough groups of distinct pairs according to their distances and computing the average for each group or tolerance region \citep[][Section 2.4]{cressie2015statistics}. Note that \cite{omre1984variogram} and \cite{reilly2007weighted} suggested to use weighted averages to obtain more resistant and efficient estimates, respectively. But in this paper, we simply use the classical point estimates, since our estimators are robust and resistant due to their shape constraints, as illustrated in Figures \ref{fig:simulation_covariance_wave} and \ref{fig:simulation_covariance_expo}. 

Now we propose an approach to inferring the covariance function $C$ with our estimators. Since we assume that $f(\mathbf{0}) = 1$ to develop our method in the regression setting, we here need to estimate the variance parameter $\sigma^2 = var (Z (\mathbf{s}) )$ as well. We suggest to estimate $C$ as follows: First, we initialize a variance estimator $\hat{\sigma}^2$ using the sample variance $\frac{1}{w} \sum_{i=1}^{w} \left( Z(\mathbf{s}_i) - \bar{Z}\right)^2 $ which is widely considered as a reasonable first estimate of $\sigma^2$  according to \citep{barnes1991variogram}. Second, we infer the function $C_0 (\mathbf{r}) = C(\mathbf{r})/\hat{\sigma}^2$ with one of the estimators proposed in this paper. Here we only use $\left( \mathbf{r}_{i,j}, \hat{c}_{i,j}\right)$ where $i\ne j$. We can update $\hat{\sigma}^2$ using the minimizer of the sum of squared differences between the point estimates and $\hat{C}(\mathbf{r}_{i, j}) =\hat{\sigma}^2 \hat{C}_0(\mathbf{r}_{i, j})$, $1 \le i, j \le w$, where $\hat{C}_0 (\mathbf{r})$ is the estimator of $C_0 (\mathbf{r})$ obtained in the second step. Then, the final covariance function estimator can be obtained by repeating the above steps.

The computational cost of our approach for covariance function estimation is proportional to the number of the point estimates, so it can be high for large point-referenced data. To reduce the computational cost, one can adopt the tolerance-region strategy mentioned above. See \citep{cressie2015statistics} for further details. 


\section{Numerical experiments \label{sec:simul}}

We conducted several simulations to demonstrate the performance of our method. We considered two different scenarios: In Section~\ref{sec:simReg}, we assumed that response $y$ was observed at $\mathbf{x}$ with additive Gaussian noise. Note that two regression equations motivated by popular covariance functions are considered. In Section~\ref{sec:simCov}, we used a stationary Gaussian process with mean zero and a wave covariance function to illustrate the efficacy of our method. Additional numerical numerical results are presented in Appendix~\ref{app:numer}.


\subsection{Regression with white noise errors \label{sec:simReg}}

\begin{figure*}[h]
    \centering
    \includegraphics[width=\textwidth]{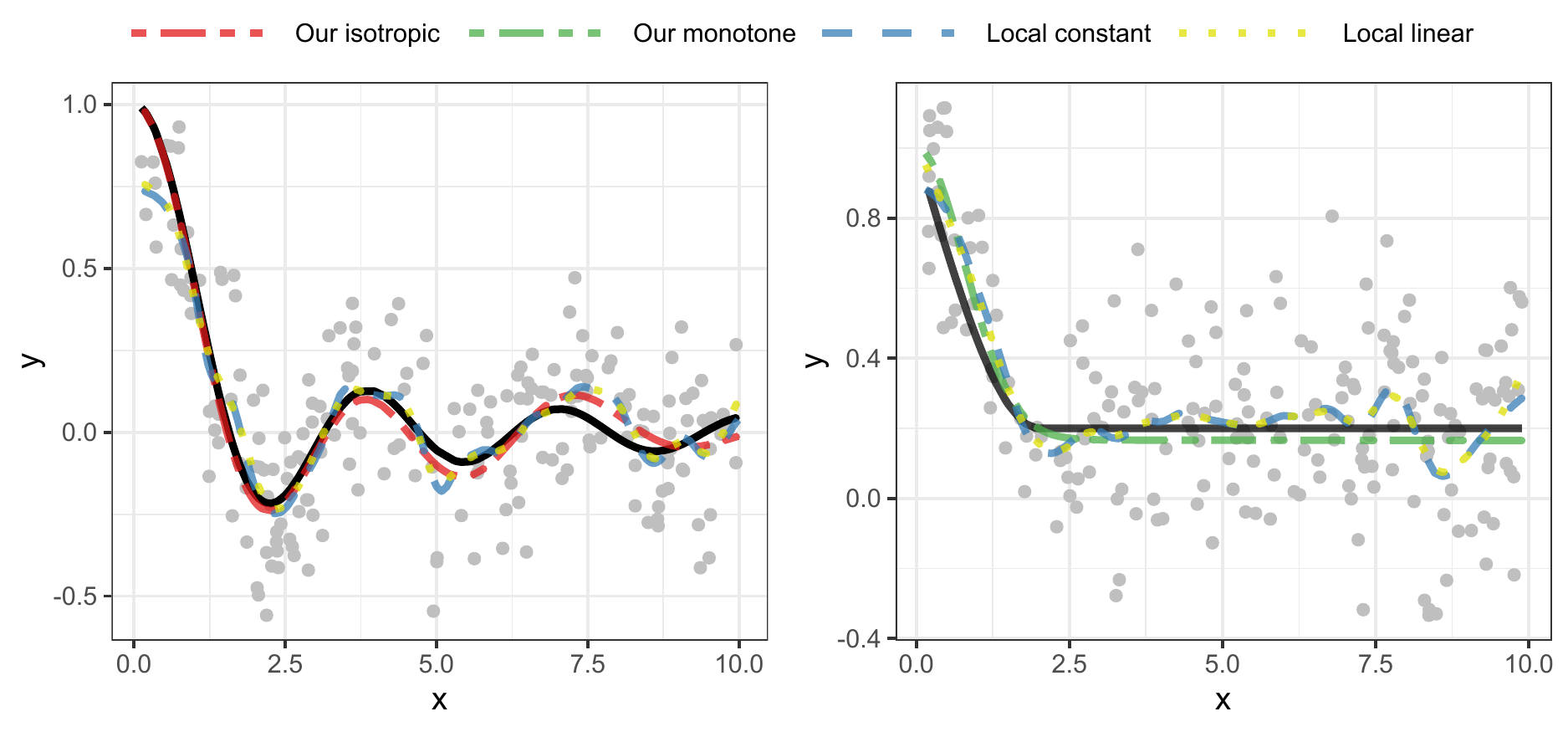}
    \caption{Comparison of our estimators to two popular nonparametric regression estimators: Local-constant and local-linear estimators. We consider the regression equations~\eqref{eq:regeqn1} and~\eqref{eq:regeqn2} and these are presented as solid black lines on the left and right panels, respectively. The gray points are their noisy observations. Our estimators are fitted using the iterated density estimation evolutionary algorithm. Bandwidth $h$ and pseudo-data size $m$ are selected using 5-fold cross validation.}
    \label{fig:simulation_regression}
\end{figure*}

We considered two different regression equations in this section. To demonstrate performance of our isotropic estimator, we considered the wave function \citep{cressie2015statistics} which is defined as follows: 
\begin{equation}
\label{eq:regeqn1}
g (r) =
\left\{
\begin{array}{ll}
	1  & \mbox{for } r = 0 \\
	\frac{\sin(2r)}{2r}  & \mbox{for } r > 0.
\end{array}
\right.
\end{equation}
On the other hand, to demonstrate performance of our monotone estimator, we considered the spherical function \citep{cressie2015statistics} which is defined as follows: 
\begin{equation}
\label{eq:regeqn2}
g (r) =
\left\{
\begin{array}{ll}
	1  & \mbox{for } r = 0 \\
	1 - \frac{1}{20} \left( 12r - r^3 \right)  & \mbox{for } 0 < r \le 2 \\
	0.2  & \mbox{for } 2 < r.
\end{array}
\right.
\end{equation}
The shapes of the functions are illustrated in Figure~\ref{fig:simulation_regression} as solid black lines. The 200 inputs are randomly selected from the interval $[0, 10]$ and the white noises are randomly drawn from the Gaussian distribution with mean of zero and standard deviation of $0.2$. As competing methods, we considered local constant (Nadaraya-Watson) and local linear estimators, since these estimators are most widely used for kernel-based regression. We used the R package \textbf{np} to obtain them \citep{hayfield2008nonparametric}. The bandwidths of the competing estimators are decided using cross validation; we use $(h, m) = (0.16, 4)$ for the isotropic estimator and $(h, m) = (0.01, 6)$ for the monotone estimator. Note that we repeated the simulation 200 times.

Our estimators outperformed the other competing estimators, as illustrated in Figure~\ref{fig:simulation_regression} and presented in Table~\ref{tbl:simulation_regression}. The plots in Figure~\ref{fig:simulation_regression} show that our estimators provided smooth curves, unlike the competing methods. Furthermore, since our approach allowed to use prior knowledge of shapes of functions including positive definiteness and monotonicity, our estimators provided better predictive performance in terms of root mean squared prediction error (RMSPE). We randomly selected test inputs from the same interval for the result presented in Table~\ref{tbl:simulation_regression}.

\begin{table}
\begin{tabular}{lcc} 
\toprule
Estimator & Wave equation & Spherical equation \\ \midrule
Local constant & 0.0533 & 0.0499 \\
Local linear & 0.0503 & 0.0430 \\
Our isotropic & 0.0419 & - \\
Our monotone & - & 0.0399 \\
\bottomrule \\
\end{tabular}
\caption{Average of the root mean squared prediction error (RMSPE) of the estimators over 200 simulated data from the regression scenario. We consider two different regression equations: For the wave equation, we use the isotropic estimator. For the spherical equation, we use the monotone estimator. These estimators are compared with local-constant and local-linear regression estimators.}
\label{tbl:simulation_regression}
\end{table}


\subsection{Covariance function estimation \label{sec:simCov}}

\begin{figure*}[p]
    \centering
    \includegraphics[width=\textwidth]{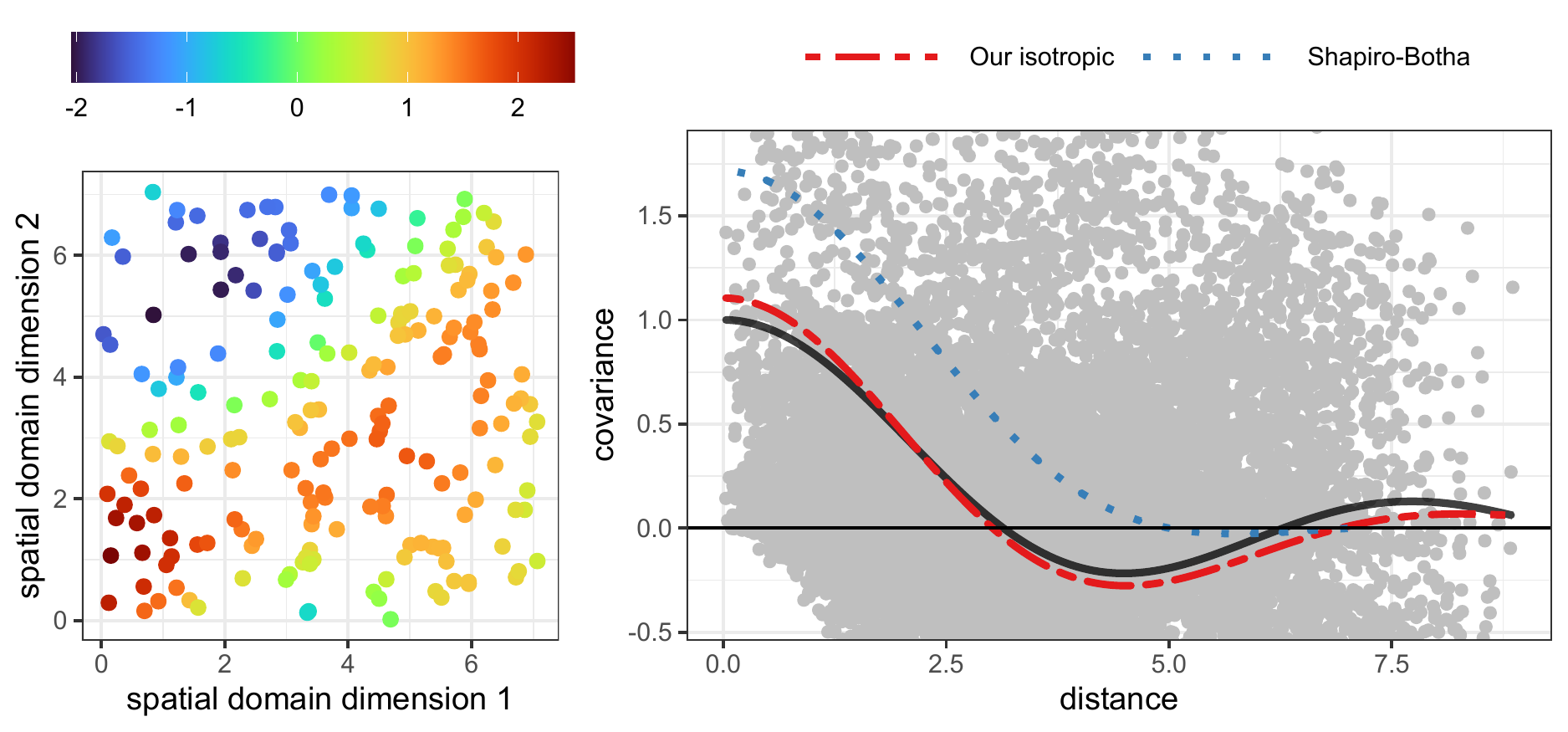}
    \caption{Comparison of our isotropic estimator to the Shapiro-Botha estimator for estimating the wave covariance function of mean-zero Gaussian process. The detailed estimation procedure is described in Section~\ref{sec:covEst}. A realization of the Gaussian process is depicted in the left panel. The true covariance function is the solid black line on the right panel. The isotropic and Shapiro-Botha estimators are two-dashed red and dotted blue lines, respectively. The gray points are point covariance estimates.}
    \label{fig:simulation_covariance_wave}
\end{figure*}

To assess the efficacy of our estimators, we considered a realization of Gaussian process with mean zero and the wave covariance function: For $\mathbf{x}$, $\mathbf{x}'$ $\in [0 , 10/\sqrt{2}]^2$, we define the wave covariance function as
\begin{equation}
\label{eq:coveqn1}
\text{cov} (\mathbf{x}, \mathbf{x}') = \frac{\sin \big(\| \mathbf{x} - \mathbf{x}' \| \big)}{\| \mathbf{x} - \mathbf{x}' \|}.
\end{equation}
A visual comparison is provided in Figure~\ref{fig:simulation_covariance_wave}. We randomly selected 200 locations in the domain so that we had 19,900 point estimates of the covariance function. Note that the competing methods considered in the previous section are not directly applicable in this setting. So, we considered the Shapiro-Botha method implemented in the R package \textbf{npsp} (0.7-8) as a competing method \citep{fernandez2019npsp}. The discretization nodes of the Shapiro-Botha method are decided following the guidelines in \cite{gorsich2004discretization}. We used cross validation to select the bandwidth $h$ and size of pseudo-datasets for our method. In specific, our isotropic estimator used $(h,m) = (0.2, 5)$. 

It appears that our method provided better performance in the setting we considered. The Shapiro-Botha estimator provided a poor estimate of the variance (or sill). Further, it did not provide covariance estimates at large distances. This is mainly because the number of data pairs whose distances are large is typically very small, so the state-of-the-art estimators can be extremely unstable. According to \cite{journel1978mining}, the maximum distance of reliability for variogram estimation is one-half the maximum distance between locations. In contrast, our method was reliable and robust even at large distances.

\section{Spatial Interpolation Comparison 97 data analysis \label{sec:realData}}

\begin{figure*}[p]
    \centering
    \includegraphics[width=\textwidth]{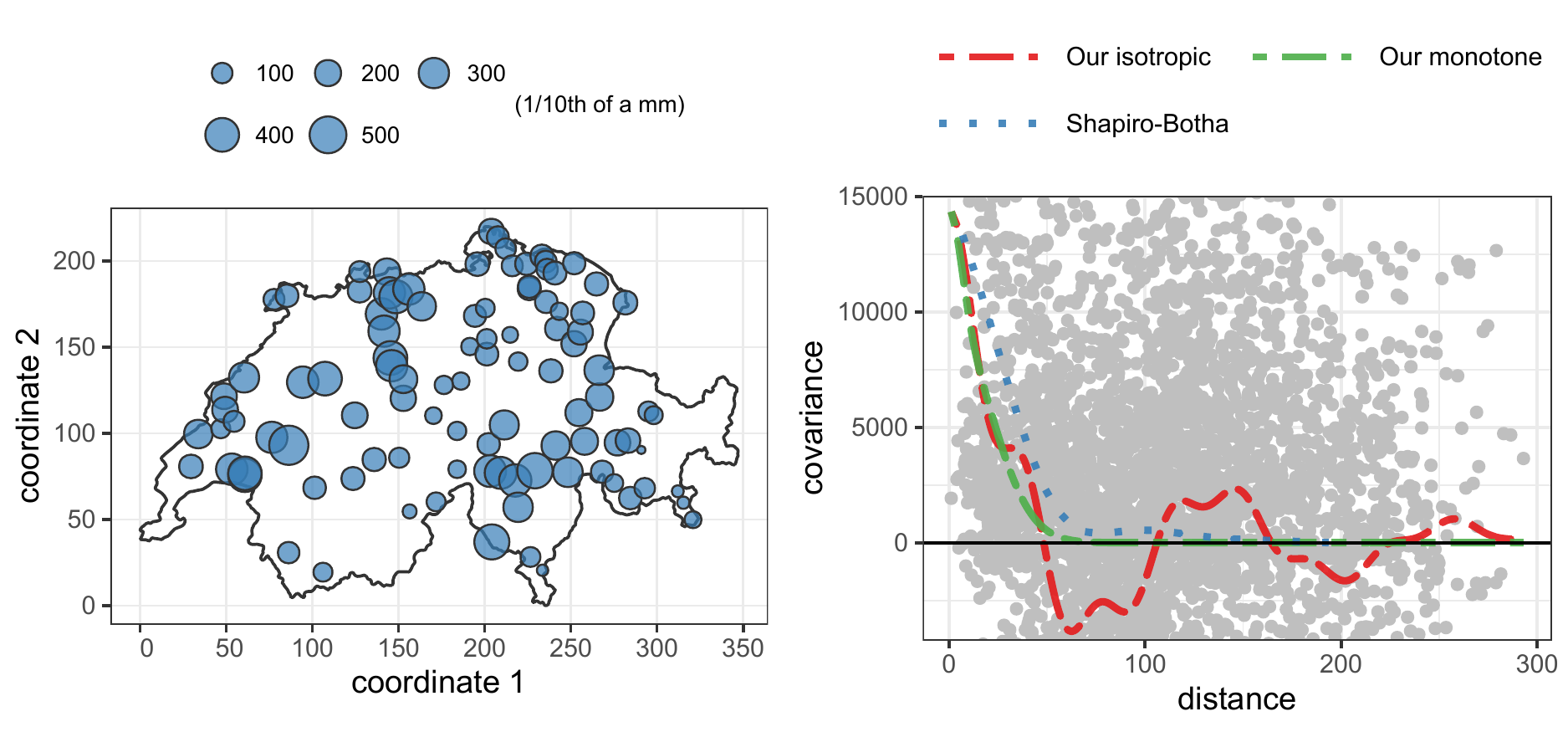}
    \caption{Proportional circle map of SIC100 data from the Spatial Interpolation Comparison (SIC) 97 project (left) and comparison of our isotropic and monotone estimators to the Shapiro-Botha estimator for covariance function estimation with SIC100 data (right). The isotropic and monotone estimators are two-dashed red and green lines, respectively, on the right panel. The Shapiro-Botha estimator is the dotted blue line. The gray points are point covariance estimates.}
    \label{fig:sic100}
\end{figure*}

We considered the data set SIC100 from the Spatial Interpolation Comparison (SIC) 97 project  \citep{dubois1998spatial} which includes 100 daily rainfall observations made in Switzerland on the 8th of May 1986 that were randomly selected from a dataset of 467 observations. A main reason we decided to use only 100 observations instead of all observations is that the project aims to estimate the rainfall at the 367 remaining locations. Furthermore, this is an ideal example to demonstrate the main advantage of our method that it provides positive definite covariance function estimators even when the sample size is very small. See \url{https://wiki.52north.org/bin/view/AI_GEOSTATS/EventsSIC97} for details. The data set can be found in the R package \textbf{geoR} \citep{diggle2007classical}. Note that we scaled distances between observations to be from 0 to 8 for numerical stability and retrieved them in Figure~\ref{fig:sic100}, and decided the bandwidth $h$ and size of pseudo data $m$ by cross validation with the initial variance estimate.

Figure~\ref{fig:sic100} shows our isotropic and monotone estimators of the underlying covariance function. Also, we again considered the Shapiro-Botha method as a competing method. Although it is not easy to make comparisons directly due to the fact that the true covariance function is unknown, it appears that our isotropic estimator is flexible enough to capture almost any underlying features of the covariance function and our monotone estimator does not suffer from spurious modes, unlike the Shapiro-Botha estimator. If there is good evidence that the covariance function is monotonically decreasing, one can simply use the monotone estimator which only needs a few minutes for hyperparameter selection using cross validation and a few tens of seconds for estimation. Note that a 64-bit workstation with 16 GB RAM and an Intel Core i7-8700K CPU running at 3.70 GHz was used. For the isotropic estimator, the bandwidth $h$ the pseudo-data size $m$ are selected to be 0.18 and 6, respectively, by cross validation. For the monotone estimator, selected $h$ and $m$ are 0.02 and 3, respectively.

\section{Discussion \label{sec:conc}}

We have introduced a new positive definite nonparametric regression method. To our understanding, this work provides a promising solution to the issue that there was no general framework to obtain positive definite nonparametric covariance function models for stationary processes. Our method allows to use popular kernels, for example, Gaussian and Epanechnikov kernels, and decide hyperparameters using cross validation, like typical nonparametric regression methods. Furthermore, our method always returns positive definite function estimators and therefore can be useful for kriging. Our numerical experiments suggested that the state-of-the-art method suggested by \cite{shapiro1991variogram} was outperformed by our estimators. But, due to the presence of heteroscedasticity for the covariance function estimation task, it was not straightforward to make fair comparisons between our method and the Shapiro-Botha method. In specific, since all the point estimates are correlated, the RMSPE cannot be a good measure to compare them.

Our approach can be easily modified for various purposes. There are several functions that can be characterized by integral transformations. For instance, completely monotone and multiply monotone functions can be expressed as integral transforms of non-negative Borel measures \citep{williamson1955multiply} so that our method can be applied for estimating these functions by using different integrands. One can use different performance measures instead of the MSE: Examples include prediction scores on hold-out observations of a spatial process and the KL divergence values of the termination step of the IDEA. We believe that further investigation is required to extend our approach.


\begin{acks}
This paper is based on a thesis submitted by the author in partial fulfillment of the requirements for a Master of Science degree from the Department of Statistics, Seoul National University. We would like to thank Matthias Katzfuss for his helpful comments.
\end{acks}

\bibliographystyle{ACM-Reference-Format}
\bibliography{additionalrefs}


\newpage
\appendix

\section{Postponed proofs \label{app:proofs}}

To derive the formulas in Section~\ref{sec:example}, we refer to \cite{gradshteyn2014table} and \cite{rosenheinrich2012tables} which provide tables of integral formulas involving the Bessel function of the first kind. In addition, for the estimators with the Gaussian kernel, we consider \cite{dattoli1993advances} which defines the generalized Bessel function and discusses its properties. Note that $\Omega_d (r) = \mathbf{J}_0 (r)$, since $d = 2$, and $\mathbf{J}_0 (-r) = \mathbf{J}_0 (r)$ for $r > 0$.

\paragraph{Uniform kernel} Let $K (u) = \frac{1}{2} \mathbf{1}_{\{ |u| \le 1 \}}$. The isotropic estimator can be written as
\begin{multline}
\hat{g}_m (r) = \frac{1}{mh} \sum_{i=1}^{m} \bigg[ \int_{0}^{\infty} \mathbf{J}_0 (ru) K \Big( \frac{u-v_i}{h} \Big) {\rm d}u \\+ \int_{0}^{\infty} \mathbf{J}_0 (ru) K \Big( \frac{u+v_i}{h} \Big) {\rm d}u\bigg].
\end{multline}
If $v_i-h \ge 0$, then
\begin{equation}
\int_{0}^{\infty} \mathbf{J}_0 (ru) K \Big( \frac{u-v_i}{h} \Big) {\rm d}u = \int_{v_i-h}^{v_i+h} \mathbf{J}_0 (ru) {\rm d}u
\end{equation}
and
\begin{equation}
\int_{0}^{\infty} \mathbf{J}_0 (ru) K \Big( \frac{u+v_i}{h} \Big) {\rm d}u = 0.
\end{equation}
From Section 6.51 of \cite{gradshteyn2014table}, one can show that
\begin{equation}
\int_{v_i-h}^{v_i+h} \mathbf{J}_0 (ru) {\rm d}u =
\frac{1}{2r} \left[ \Lambda_1 (r(v_j +h)) - \Lambda_1 (r(v_j - h)) \right].
\end{equation}
If $v_i-h < 0$, then
\begin{equation}
\int_{0}^{\infty} \mathbf{J}_0 (ru) K \Big( \frac{u-v_i}{h} \Big) {\rm d}u = \int_{0}^{v_i+h} \mathbf{J}_0 (ru) {\rm d}u
\end{equation}
and
\begin{equation}
\int_{0}^{\infty} \mathbf{J}_0 (ru) K \Big( \frac{u+v_i}{h} \Big) {\rm d}u = \int_{0}^{-v_i+h} \mathbf{J}_0 (ru) {\rm d}u.
\end{equation}
From Section 6.51 of \cite{gradshteyn2014table} and the fact that $\mathbf{J}_0$ is even, one can show that
\begin{multline}
\int_{0}^{v_i+h} \mathbf{J}_0 (ru) {\rm d}u + \int_{0}^{-v_i+h} \mathbf{J}_0 (ru) {\rm d}u \\ = \frac{1}{2r} \left[ \Lambda_1 (r(v_j +h)) - \Lambda_1 (r(v_j - h)) \right].
\end{multline}
Therefore, 
\begin{equation}
\hat{g}_m (r) = \frac{1}{2mhr} \sum_{j=1}^{m} \left[ \Lambda_1 (r(v_j +h)) - \Lambda_1 (r(v_j - h)) \right].
\end{equation}
The monotone estimator can be written as
\begin{multline}
\hat{q}_m (r) = \frac{1}{mh} \sum_{i=1}^{m} \bigg[ \int_{0}^{\infty} e^{-r^2 u^2} K \Big( \frac{u-v_i}{h} \Big) {\rm d}u \\+ \int_{0}^{\infty} e^{-r^2 u^2} K \Big( \frac{u+v_i}{h} \Big) {\rm d}u\bigg].
\end{multline}
If $v_i-h \ge 0$, then
\begin{equation}
\int_{0}^{\infty} e^{-r^2 u^2} K \Big( \frac{u-v_i}{h} \Big) {\rm d}u = \frac{1}{\sqrt{2} r} \int_{\sqrt{2} r (v_i-h)}^{\sqrt{2} r (v_i+h)} e^{-t^2 / 2} {\rm d}t
\end{equation}
and
\begin{equation}
\int_{0}^{\infty} e^{-r^2 u^2} K \Big( \frac{u+v_i}{h} \Big) {\rm d}u = 0.   
\end{equation}
From the definition of the standard Gaussian distribution,
\begin{multline}
\int_{0}^{\infty} e^{-r^2 u^2} K \Big( \frac{u-v_i}{h} \Big) {\rm d}u \\ = \frac{\sqrt{\pi}}{r} \left[ \Phi \left(\sqrt{2}r(v_j +h)\right) - \Phi \left(\sqrt{2}r(v_j -h)\right) \right].
\end{multline}
If $v_i-h < 0$, then
\begin{equation}
\int_{0}^{\infty} e^{-r^2 u^2} K \Big( \frac{u-v_i}{h} \Big) {\rm d}u = \frac{1}{\sqrt{2} r} \int_{0}^{\sqrt{2} r (v_i+h)} e^{-t^2 / 2} {\rm d}t
\end{equation}
and
\begin{equation}
\int_{0}^{\infty} e^{-r^2 u^2} K \Big( \frac{u+v_i}{h} \Big) {\rm d}u = \frac{1}{\sqrt{2} r} \int_{0}^{\sqrt{2} r (-v_i+h)} e^{-t^2 / 2} {\rm d}t.   
\end{equation}
From the fact that $e^{-t^2 / 2}$ is even, one can similarly see that 
\begin{multline}
\frac{1}{\sqrt{2} r} \int_{0}^{\sqrt{2} r (v_i+h)} e^{-t^2 / 2} {\rm d}t + \frac{1}{\sqrt{2} r} \int_{0}^{\sqrt{2} r (-v_i+h)} e^{-t^2 / 2} {\rm d}t \\ = \frac{\sqrt{\pi}}{r} \left[ \Phi \left(\sqrt{2}r(v_j +h)\right) - \Phi \left(\sqrt{2}r(v_j -h)\right) \right].
\end{multline}
Therefore,
\begin{equation}
\hat{q}_m (r) = \frac{\sqrt{\pi}}{2mhr} \sum_{j = 1}^{m} \left[ \Phi \left(\sqrt{2}r(v_j +h)\right) - \Phi \left(\sqrt{2}r(v_j -h)\right) \right],
\end{equation}

\paragraph{Epanechnikov kernel} Let $K (u) = \frac{3}{4} \left( 1 - u^2\right) \mathbf{1}_{\{ |u| \le 1 \}}$. The proof is similar to the previous one for the uniform kernel case, but in addition we need the following facts:
\begin{equation}
\int u \mathbf{J}_0 (u) {\rm d}u = u \mathbf{J}_1 (u)
\end{equation}
up to the constant of integration and
\begin{equation}
\int u^2 \mathbf{J}_0 (u) {\rm d}u = u^2 \mathbf{J}_1 (u) + u \mathbf{J}_0 (u) - \int \mathbf{J}_0 (u) {\rm d}u.
\end{equation}

\paragraph{Gaussian kernel} Let $K (u) = \frac{1}{\sqrt{2 \pi}} e^{-\frac{1}{2}u^2}$. Note that $\mathbf{J}_{l}(-a) = (-1)^l \mathbf{J}_{l}(a)$, $\mathbf{J}_{-l}(a) = (-1)^l \mathbf{J}_{l}(a)$ and $$\mathbf{J}_{0}(a+b) = \sum_{l = -\infty}^{\infty} \mathbf{J}_{l} (a) \mathbf{J}_{-l} (b)$$ for integer order $l$. From Section 6.618 of \cite{gradshteyn2014table},
\begin{align}
\hat{g}_m (r) &= \frac{1}{m} \sum_{i=1}^{m} \bigg[ \frac{2}{\sqrt{2\pi}} \sum_{l = -\infty}^{\infty} \mathbf{J}_{2l} (r v_i) \int_{0}^{\infty} e^{-\frac{1}{2}t^2} \mathbf{J}_{2l} (hrt) {\rm d}t \bigg] \\
&= \frac{1}{m} e^{-\frac{1}{4} h^2 r^2} \sum_{j=1}^{m} \bigg[ \sum_{l = -\infty}^{\infty} \mathrm{i}^{-l} \mathbf{J}_{l} \Big(\frac{\mathrm{i}}{4} h^2 r^2 \Big) \mathbf{J}_{2l} (r v_i)\bigg] \\
&= \frac{1}{m} e^{-\frac{1}{4} h^2 r^2} \sum_{j=1}^{m} \mathbf{J}_0 ^{(2)} \left(\frac{\mathrm{i}\mkern1mu}{4}h^2 r^2 , rv_j , {\mathrm{i}\mkern1mu}^{-1}\right)
\end{align}
where the $\myatop{(2)}{0}$ order generalized Bessel function of the first kind $\mathbf{J}_0 ^{(2)}$ is defined by \cite{dattoli1993advances}. Furthermore, the integral representation of $\hat{g}_m (r)$ can be obtained using that of $\mathbf{J}_0 ^{(2)}$ presented in \cite{dattoli1993advances}.


\section{Additional numerical results \label{app:numer}}

\begin{figure*}
    \centering
    \includegraphics[width=\textwidth]{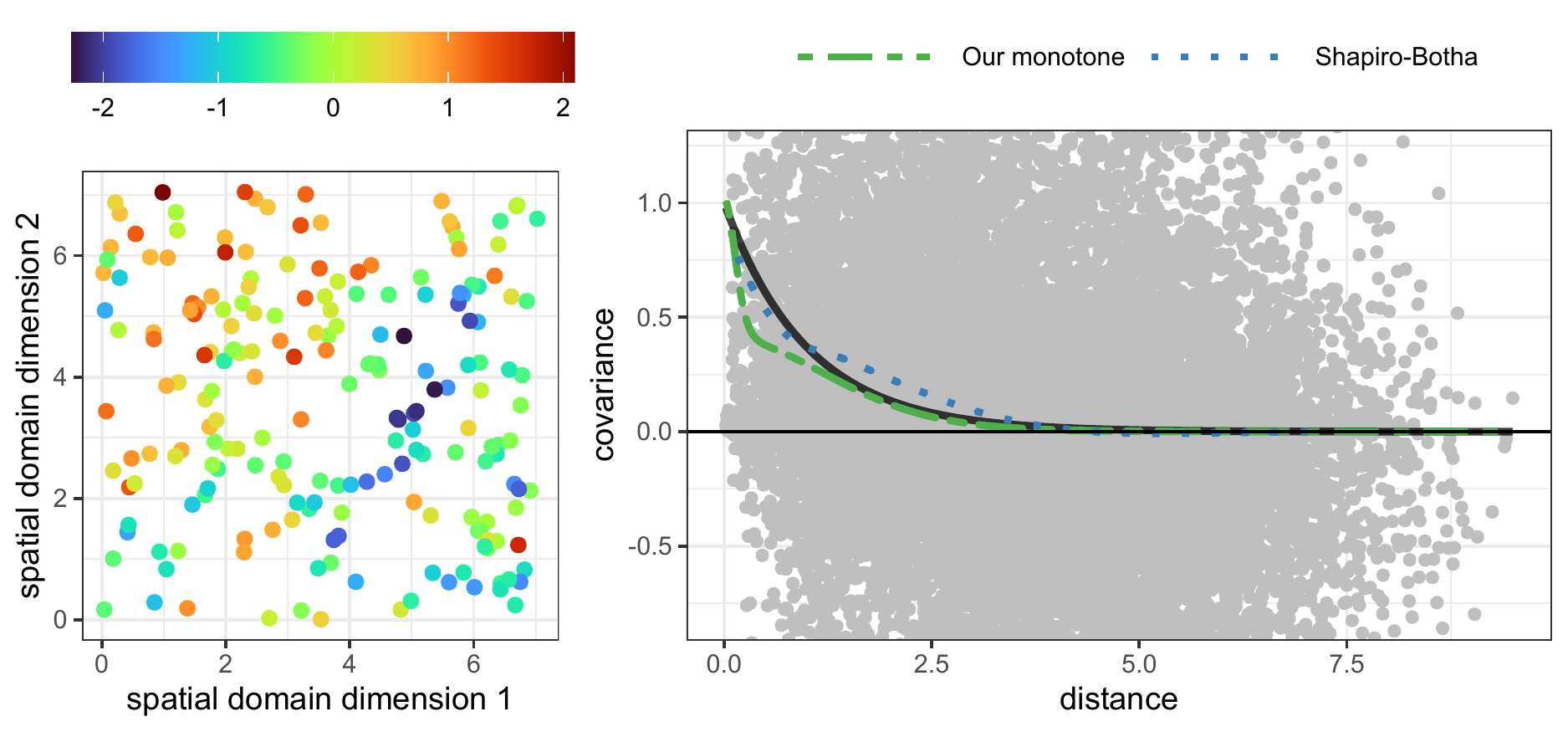}
    \caption{Comparison of our isotropic estimator to the Shapiro-Botha estimator for estimating the exponential covariance function of mean-zero Gaussian process. The detailed estimation procedure is described in Section~\ref{sec:covEst}. A realization of the Gaussian process is depicted in the left panel. The true covariance function is the solid black line on the right panel. The isotropic and Shapiro-Botha estimators are two-dashed red and dotted blue lines, respectively. The gray points are point covariance estimates.}
    \label{fig:simulation_covariance_expo}
\end{figure*}

We considered a realization of a Gaussian process with mean zero and exponential covariance function: For $\mathbf{x}$, $\mathbf{x}'$ $\in [0 , 10/\sqrt{2}]^2$, we define the exponential covariance function as
\begin{equation}
\label{eq:coveqn2}
\text{cov} (\mathbf{x}, \mathbf{x}') = \exp \big(- \| \mathbf{x} - \mathbf{x}' \| \big).
\end{equation}
A visual comparison is provided in Figure~\ref{fig:simulation_covariance_expo}. We randomly selected 200 locations in the domain so that we had 19,900 point estimates of the covariance function. we considered the Shapiro-Botha method implemented in the R package \textbf{npsp} (0.7-8) as a competing method \citep{fernandez2019npsp}. The discretization nodes of the Shapiro-Botha method are decided following the guideline in \cite{gorsich2004discretization}. We used cross validation to select the bandwidth $h$ and the size of pseudo-datasets for our method. In specific, our monotone estimator used $(h,m) = (0.1,10)$.


\section{Convergence of the algorithm \label{app:conv}}

We here provide empirical evidence on the convergence of the iterated density estimation evolutionary algorithm (IDEA) in several regression settings. We consider three regression equations:
\begin{enumerate}[leftmargin=0.75cm]
    \item[$\cdot$] Wave function with a scale $c >0$ : $$g (r) =
\left\{
	\begin{array}{ll}
		1  & \mbox{for } r = 0 \\
		\frac{\sin(cr)}{cr}  & \mbox{for } r > 0,
	\end{array}
\right.$$
    \item[$\cdot$] Spherical function with a scale $c > 0$ and support $[0, b]$: $$g (r) =
\left\{
	\begin{array}{ll}
		1  & \mbox{for } r = 0 \\
		1 - \frac{b}{2} \left( 3cr - c^3 r^3 \right)  & \mbox{for } 0 < r \le c^{-1} \\
		1 - b  & \mbox{for } c^{-1} < r,
	\end{array}
\right.$$
    \item[$\cdot$] Exponential function with a scale $c>0$: $$g (r) = \exp \left(- cr \right) \quad \mbox{for } r > 0.$$
\end{enumerate}
Note that these functions are widely used isotropic covariance models \citep{cressie2015statistics}. For the wave model, we consider two different scale parameter values: The wave functions with $c=1$ and $c=2$ are called small-scale and large-scale wave functions, respectively. For the Spherical model, we consider two different settings: The spherical functions with $(b,c) = (1, 1)$ and $(b,c) = (0.8, 0.5)$ are called narrowly and widely supported spherical functions, respectively. For the exponential model, we consider two different scale parameter values: The exponential functions with $c=1$ and $c=0.25$ are called small scale and large-scale exponential functions, respectively. For simplicity, the 200 values of $r$ are randomly sampled from the interval $[0, 10]$ and the observations are contaminated by the white noise $\varepsilon \sim \mathcal{N} (0, 0.2^2)$.

\begin{figure*}[p]
    \centering
    \includegraphics[width=0.95\textwidth]{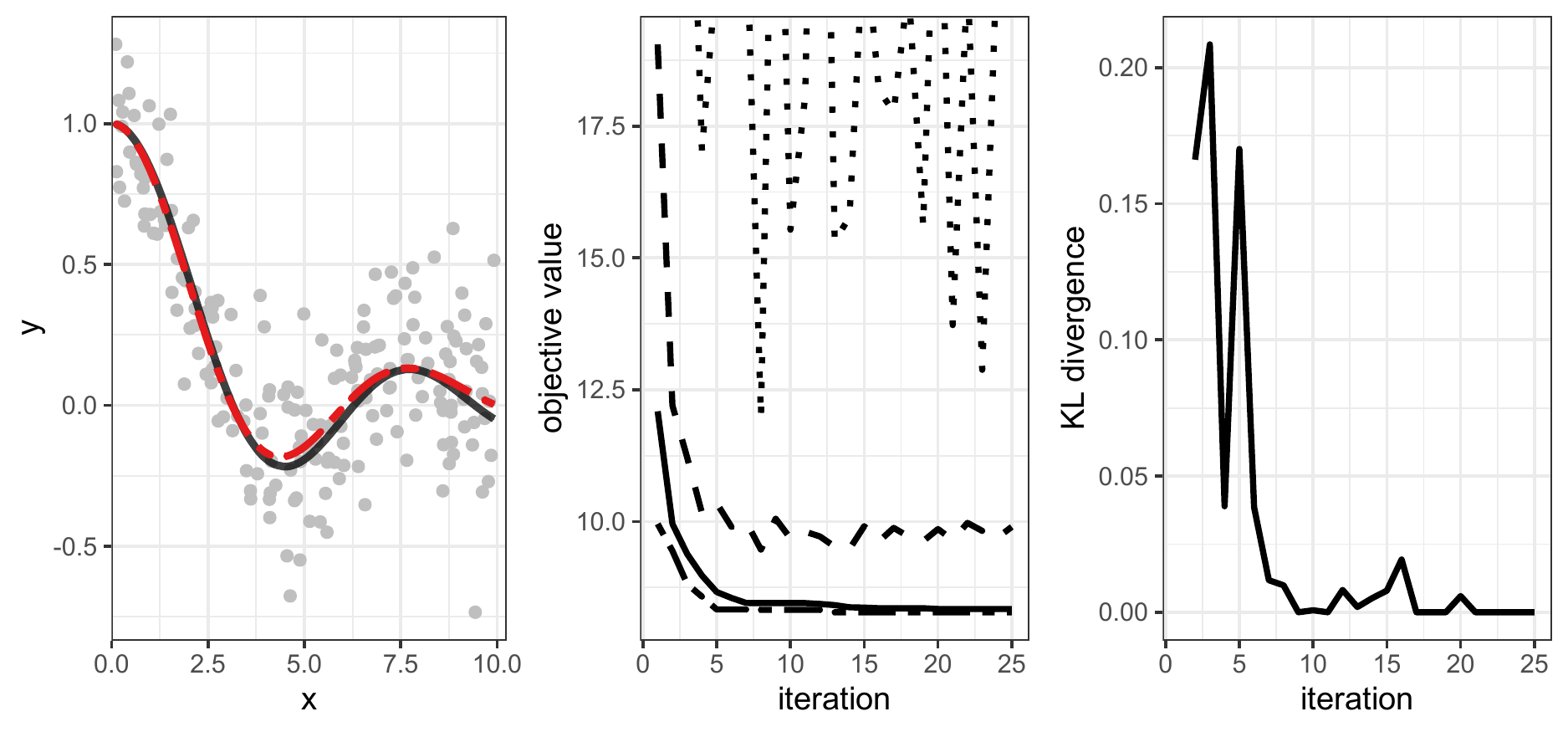}
    \caption{Convergence of the iterated density estimation evolutionary algorithm (IDEA) for fitting the isotropic estimator to the small-scale wave regression equation: The left panel shows the true regression equation (solid black line), its isotropic estimator with Gaussian kernel (two-dashed red line), and cloud of noisy observations (gray points). The center panel shows the trace plot of the objective functions values across iteration of IDEA. The lines represent, from below, the minimum (two-dashed), largest selected (solid), mean (dashed), and maximum (dotted) objective function values of pseudo datasets at iterations. The right panel shows the trace plot of the KL divergence defined for the termination step.}
    \label{fig:simulation_convergence_wave_simple}
\end{figure*}

\begin{figure*}[p]
    \centering
    \includegraphics[width=0.95\textwidth]{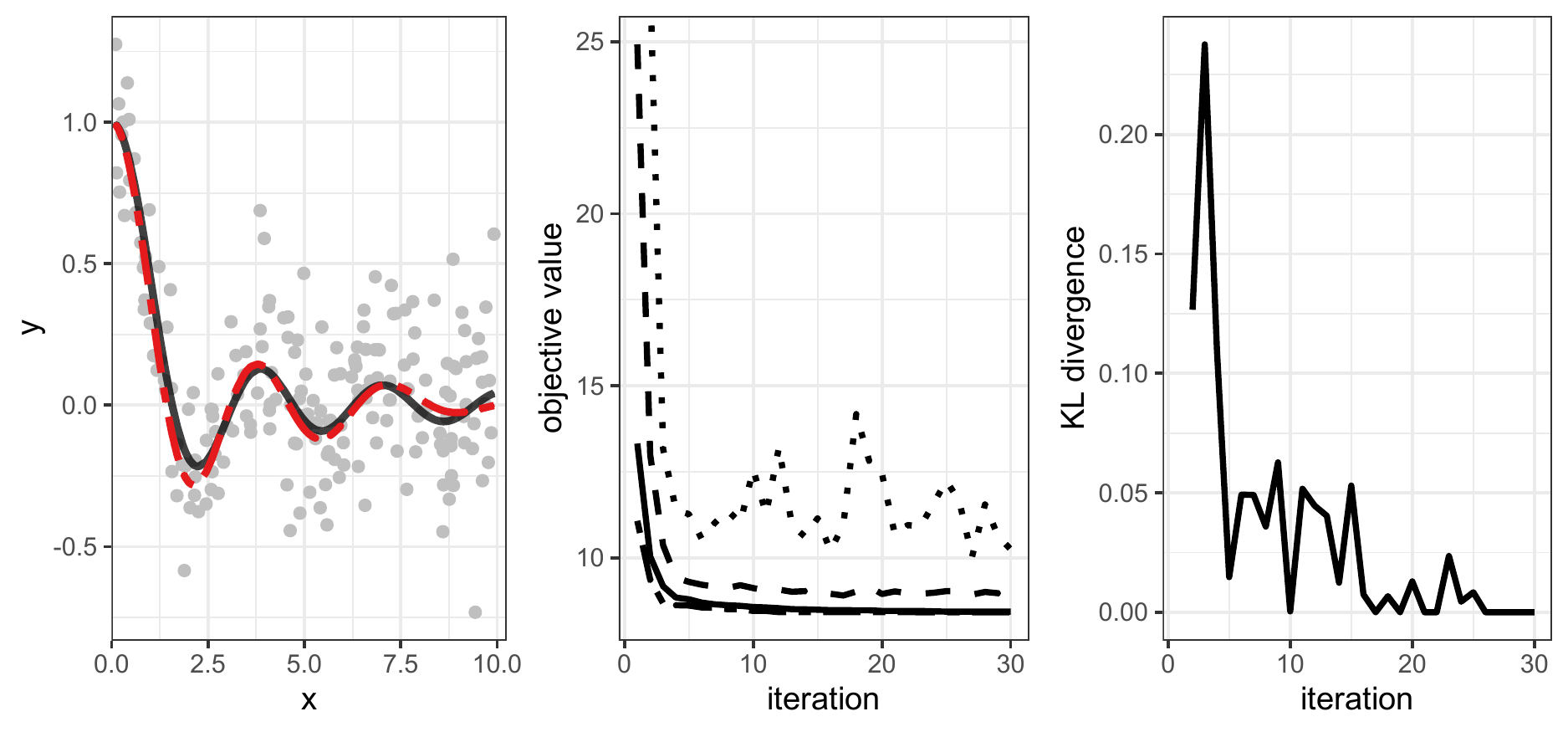}
    \caption{Convergence of the iterated density estimation evolutionary algorithm (IDEA) for fitting the isotropic estimator to the large-scale wave regression equation: The left panel shows the true regression equation (solid black line), its isotropic estimator with Gaussian kernel (two-dashed red line), and cloud of noisy observations (gray points). The center panel shows the trace plot of the objective functions values across iteration of IDEA. The lines represent, from below, the minimum (two-dashed), largest selected (solid), mean (dashed), and maximum (dotted) objective function values of pseudo datasets at iterations. The right panel shows the trace plot of the KL divergence defined for the termination step.}
    \label{fig:simulation_convergence_wave_complex}
\end{figure*}

\begin{figure*}[p]
    \centering
    \includegraphics[width=0.95\textwidth]{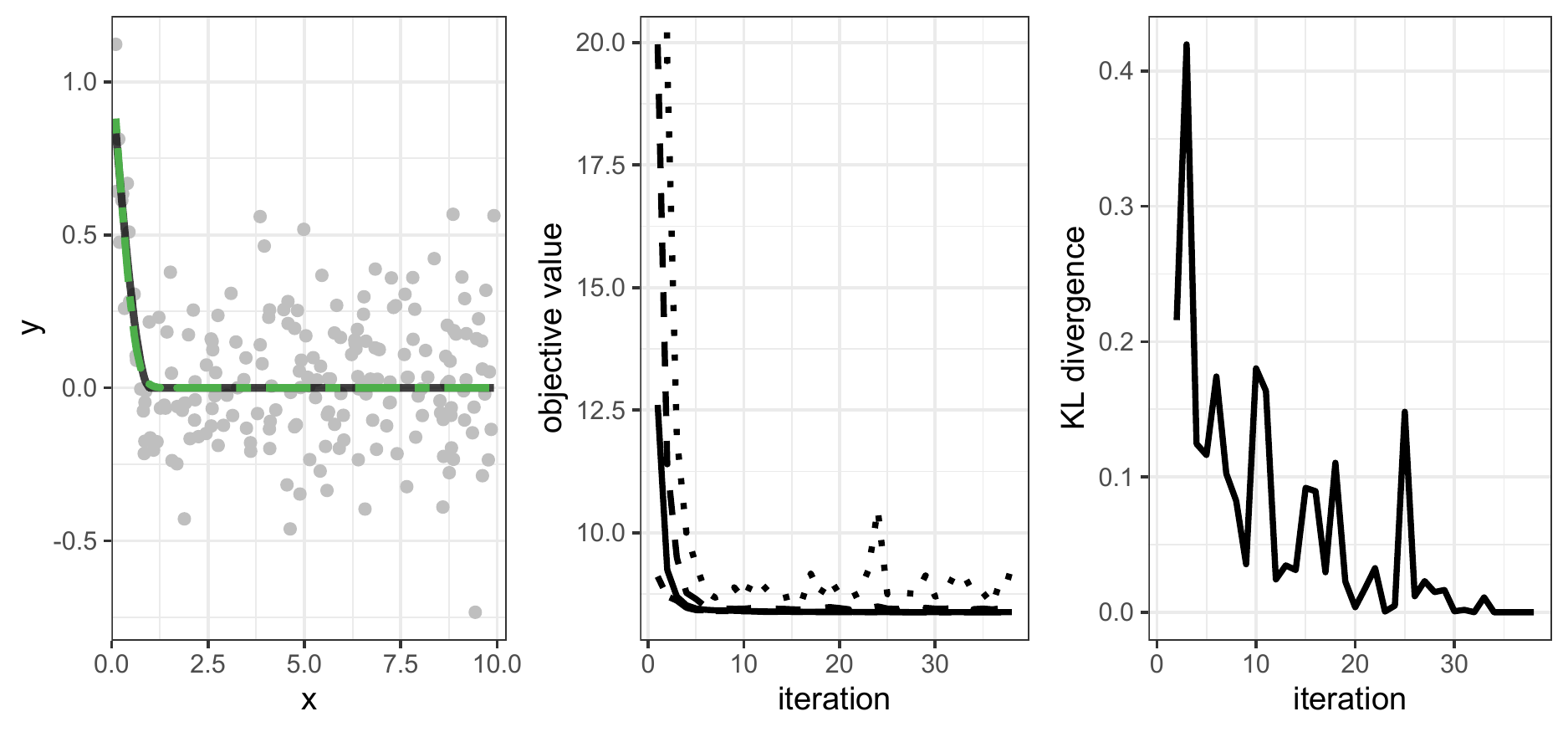}
    \caption{Convergence of the iterated density estimation evolutionary algorithm (IDEA) for fitting the monotone estimator to the narrowly supported spherical regression equation: The left panel shows the true regression equation (solid black line), its monotone estimator with Gaussian kernel (two-dashed green line), and cloud of noisy observations (gray points). The center panel shows the trace plot of the objective functions values across iteration of IDEA. The lines represent, from below, the minimum (two-dashed), largest selected (solid), mean (dashed), and maximum (dotted) objective function values of pseudo datasets at iterations. The right panel shows the trace plot of the KL divergence defined for the termination step.}
    \label{fig:simulation_convergence_sphe_simple}
\end{figure*}

\begin{figure*}[p]
    \centering
    \includegraphics[width=0.95\textwidth]{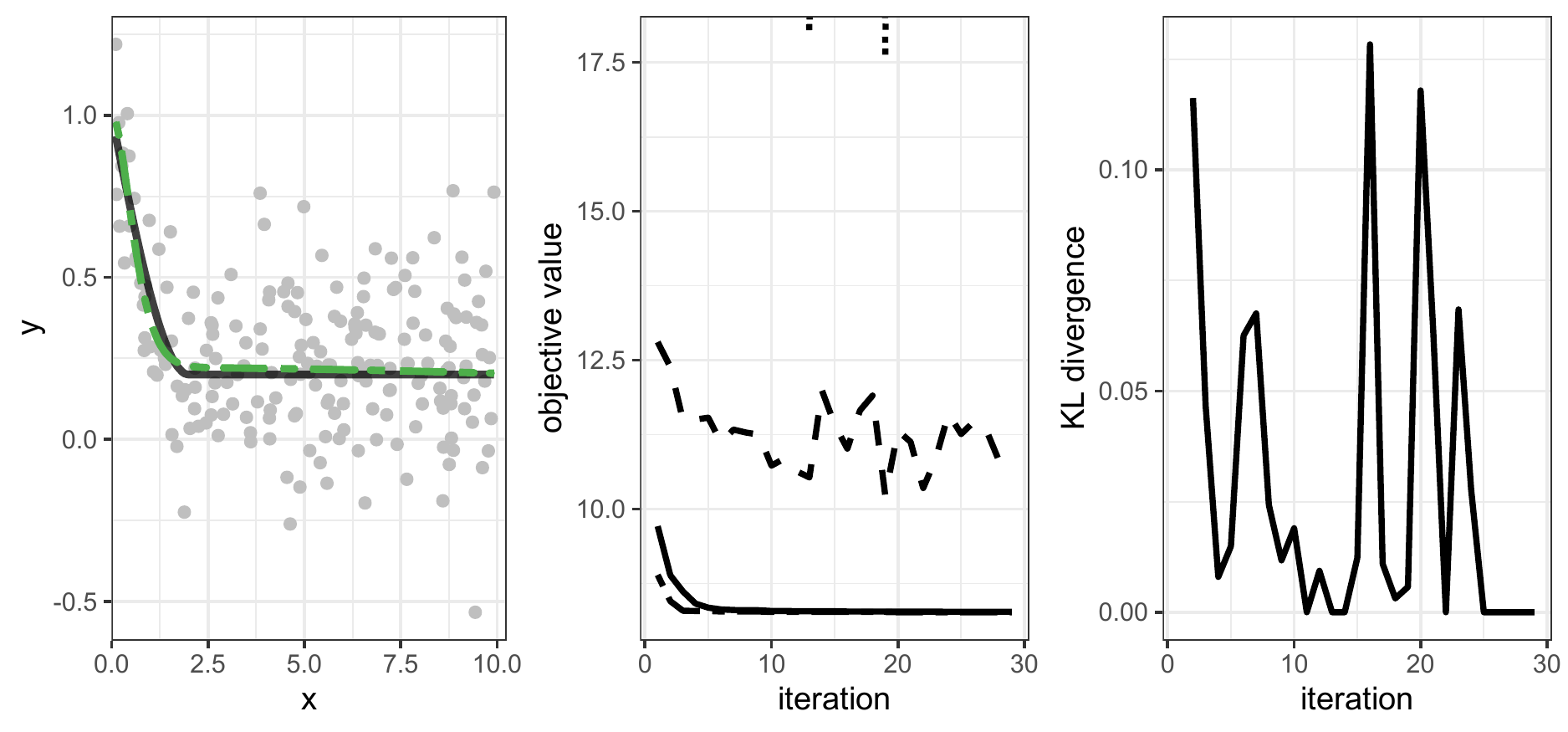}
    \caption{Convergence of the iterated density estimation evolutionary algorithm (IDEA) for fitting the monotone estimator to the widely supported spherical regression equation: The left panel shows the true regression equation (solid black line), its monotone estimator with Gaussian kernel (two-dashed green line), and cloud of noisy observations (gray points). The center panel shows the trace plot of the objective functions values across iteration of IDEA. The lines represent, from below, the minimum (two-dashed), largest selected (solid), mean (dashed), and maximum (dotted) objective function values of pseudo datasets at iterations. The right panel shows the trace plot of the KL divergence defined for the termination step.}
    \label{fig:simulation_convergence_sphe_complex}
\end{figure*}

\begin{figure*}[p]
    \centering
    \includegraphics[width=0.95\textwidth]{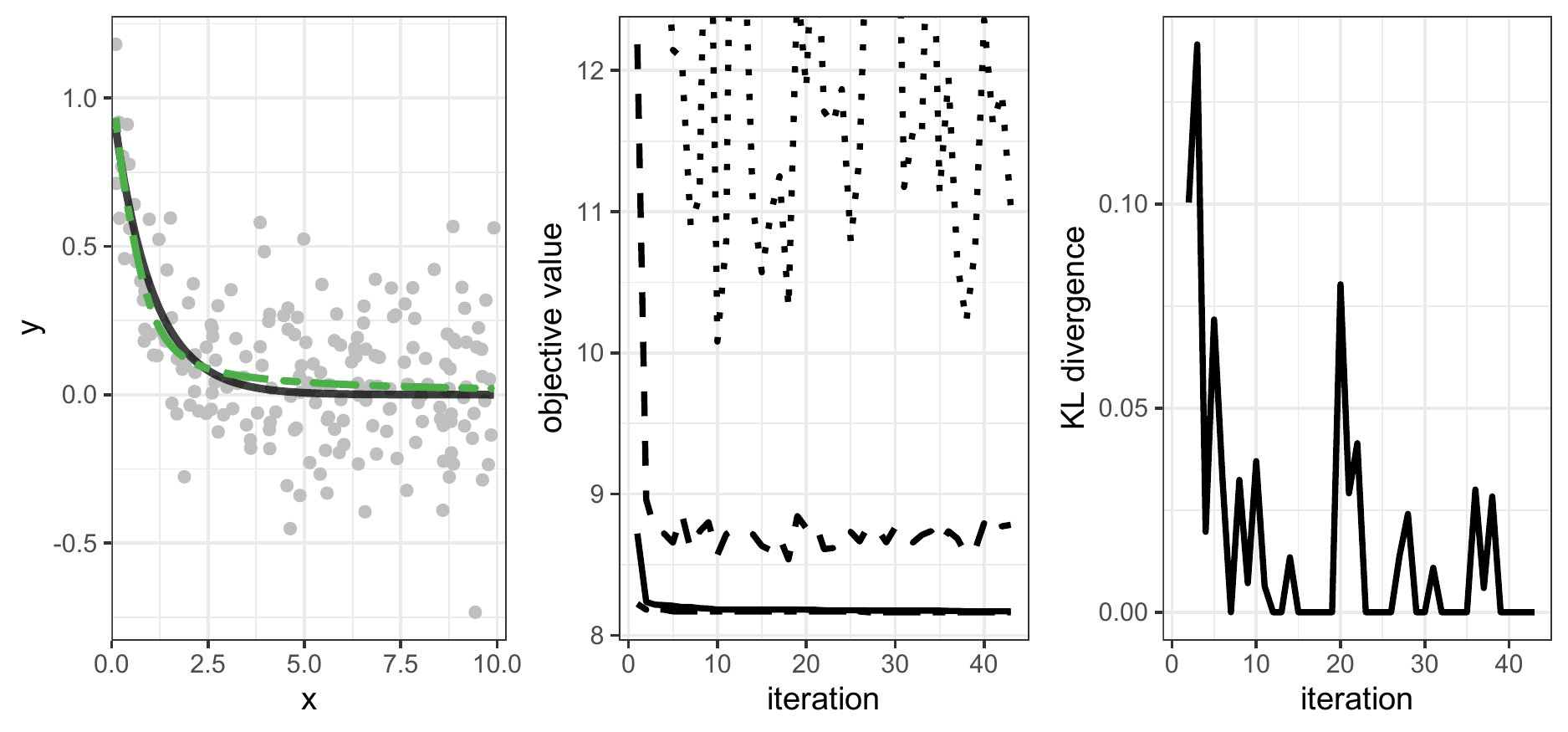}
    \caption{Convergence of the iterated density estimation evolutionary algorithm (IDEA) for fitting the monotone estimator to the small-scale exponential regression equation: The left panel shows the true regression equation (solid black line), its monotone estimator with Gaussian kernel (two-dashed green line), and cloud of noisy observations (gray points). The center panel shows the trace plot of the objective functions values across iteration of IDEA. The lines represent, from below, the minimum (two-dashed), largest selected (solid), mean (dashed), and maximum (dotted) objective function values of pseudo datasets at iterations. The right panel shows the trace plot of the KL divergence defined for the termination step.}
    \label{fig:simulation_convergence_expo_simple}
\end{figure*}

\begin{figure*}[p]
    \centering
    \includegraphics[width=0.95\textwidth]{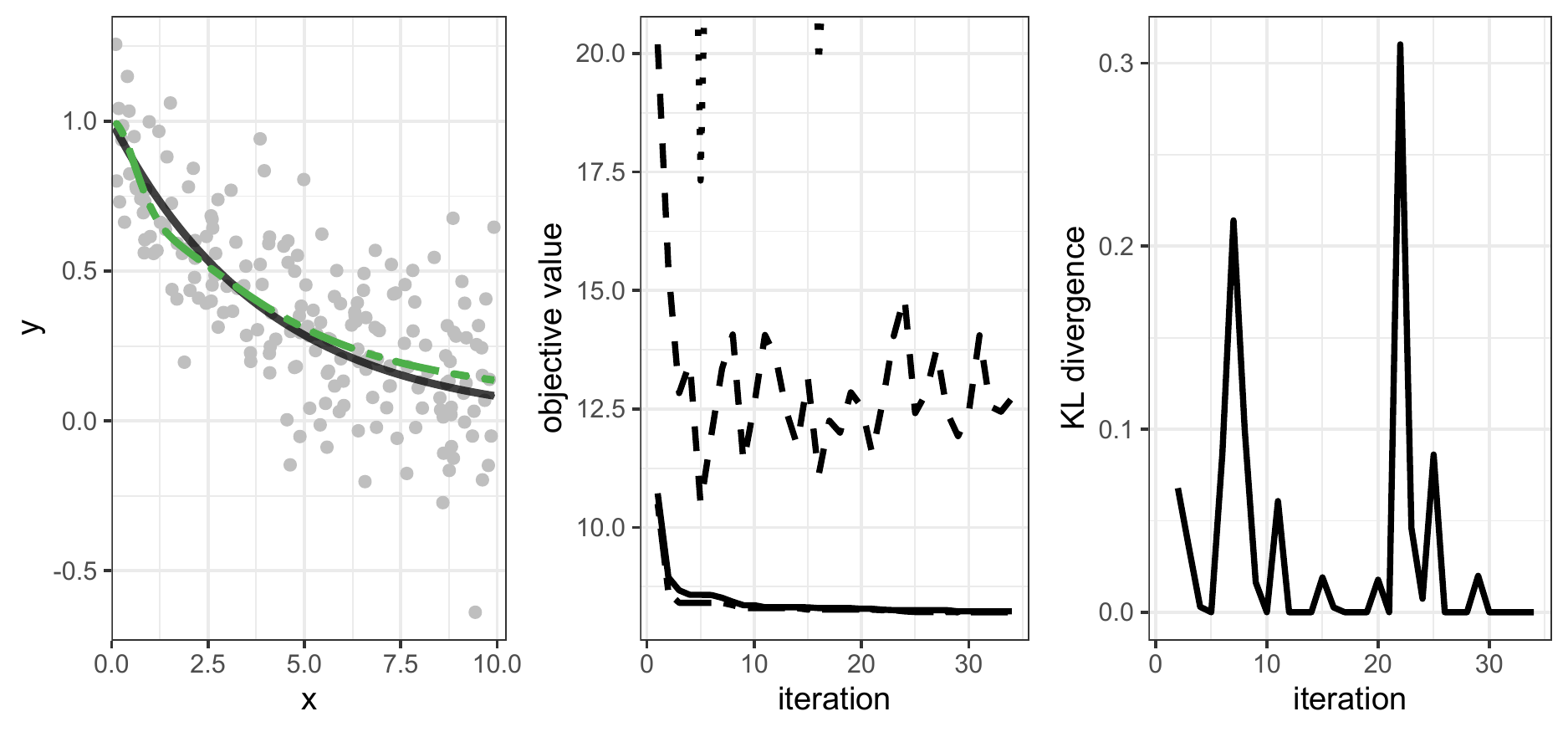}
    \caption{Convergence of the iterated density estimation evolutionary algorithm (IDEA) for fitting the monotone estimator to the large-scale exponential regression equation: The left panel shows the true regression equation (solid black line), its monotone estimator with Gaussian kernel (two-dashed green line), and cloud of noisy observations (gray points). The center panel shows the trace plot of the objective functions values across iteration of IDEA. The lines represent, from below, the minimum (two-dashed), largest selected (solid), mean (dashed), and maximum (dotted) objective function values of pseudo datasets at iterations. The right panel shows the trace plot of the KL divergence defined for the termination step.}
    \label{fig:simulation_convergence_expo_complex}
\end{figure*}

The figures~\ref{fig:simulation_convergence_wave_simple}-\ref{fig:simulation_convergence_expo_complex} in this section demonstrate estimating results and trace plots of the IDEA. For the wave models, we used the isotropic model. For the others, we used the monotone model. Since the models allow users to utilize that fact that the function value at zero is equal to one and the shapes of the functions satisfy positive definiteness, the estimators perform very well. Also, according to the trace plots, the IDEA terminates does not require more than 50 iterations in all the scenarios we considered here.

\end{document}